\documentstyle[11pt,newpasp,twoside,epsf]{article}
\def\msol{{\rm\,M_\odot}}
\def\kms{{\rm\,km\,s^{-1}}}

\def\edcomment#1{\iffalse\marginpar{\raggedright\sl#1\/}\else\relax\fi}
\reversemarginpar

\begin{document}
\title{Enrichment of the High-Redshift IGM by Galactic Winds}
 \author{Anthony Aguirre, Joop Schaye}
\affil{School of Natural Sciences, Institute for Advanced Study\\
Princeton, NJ 08540}
\author{Lars Hernquist}
\affil{Department of Astronomy, Harvard University\\
 60 Garden St., Cambridge, MA 02138}
\author{David H. Weinberg}
\affil{Department of Astronomy, Ohio State University\\
 Columbus, OH 43210}
\author{Neal Katz}
\affil{Department of Astronomy, University of Massachusetts\\
 Amherst, MA 01003}
\author{Jeffrey Gardner}
\affil{Department of Physics and Astronomy, University of Pittsburgh\\
 Pittsburgh, PA 15260}

\begin{abstract}
This paper discusses a semi-numerical method of investigating the
enrichment of the intergalactic medium by galactic winds.  We find
that most galaxies at $z \ga 3$ should be driving winds, and that (if
these winds are similar to those at low $z$) these winds should escape
to large distances.  Our calculations -- which permit exploration of a
large region of model parameter space -- indicate that the wind
velocity, the mass of the wind-driving galaxies, the fraction of
ambient material entrained, and the available time (between wind
launch and the observed redshift) all affect wind propagation
significantly; other physical effects can be important but are
sub-dominant. We find that under reasonable assumptions, the enrichment
by $3 \la z \la 6$ galaxies could account for the quantity of metals
seen in the Ly$\alpha$ forest, though it is presently unclear whether
this enrichment is compatible with the intergalactic medium's detailed
metal distribution or relative quiescence.
\end{abstract}

\section{Introduction}

	Although in the standard hot big-bang cosmology all heavy
elements (`metals') form in stars, which are in turn expected to form
in galaxies, the intergalactic medium (IGM) from which galaxies form
is nevertheless enriched with metals, at all redshifts and densities
yet observed.  Even in rather low density (overdensity $\delta \la 5$)
regions, the IGM appears to be polluted to $\ga 0.1\%$ solar
metallicity, as demonstrated by metal lines in Ly$\alpha$ absorption
systems (e.g., Cowie et al. 1995).  Exactly
how this enrichment occurred constitutes one of the most interesting
unsolved problems in the study of galaxy and structure formation.
Possible enrichment mechanisms which have been discussed in the
literature are the dynamical removal of metal-rich gas (in
ram-pressure stripping or through galaxy interactions), the ejection
of dust by radiation pressure, and supernova-driven galactic winds.

	Aguirre et al. (2001a) have recently presented a method of
invesigating enrichment due to all three mechanisms, using a hybrid
numerical/analytical approach in which metals are added to
already-completed cosmological hydrodynamic simulations in a way that
approximates their distribution by each mechanism.  This method has
been used to study the pollution of the IGM by dust (Aguirre et
al. 2001c) and the enrichment of the low-density IGM at $z \ga 3$ by
winds (Aguirre et al. 2001b).  Here I focus on the latter topic.

\section{Calculating Wind Enrichment}

	Supernova-driven `superwinds' develop in galaxies when the
supernova rate per unit volume is high enough that supernova remnants
overlap before they cool.  This leads to a single `superbubble' which,
if it can break out of the disk, flows into the IGM as a
galactic-scale wind.  Such outflows are observed nearby in starburst
galaxies (e.g., Heckman, Armus, \& Miley 1990), and inferred at
$z\sim3$ from the spectra of Lyman-break galaxies (e.g., Pettini et
al. 2001).  What these observations do {\em not} tell us is how far
into the IGM the winds may propagate (though there are some
indications; see below), and to what degree (and with what variation)
they enrich the IGM.

	To calculate such things using our method we start with a
number of outputs from a cosmological simulation incorporating star
formation.  At each step, we determine which stars have formed since
the last step, and assume that they instantaneously generate some
yield of metals.  These metals are then distributed to neighboring gas
particles either `locally', or via a prescription to simulate their
dispersal by winds.  The accumulated enrichment of the IGM at lower
redshifts can then be tracked.  In the wind dispersal prescription
(see Aguirre et al. 2001a for details), galaxies with a
star formation rate per unit area exceeding SFR$_{\rm crit} =
0.1\msol{\rm yr^{-1}\,kpc^{-2}}$ are assumed to drive steady-state,
mass-conserving winds with velocity $v_{\rm out}$ at the center of
star formation, and with mass outflow rate of $1.3\chi$ times their
SFR (with $\chi\sim1$).  The assumed supernova efficiency ($\chi$),
critical SFR, and outflow velocity are based on observations of winds
in starburst and Lyman-break galaxies.  For each of a
number $N$ of angles proportional to the galaxy's mass, a `test shell'
is propagated under the forces of gravity, the wind ram pressure, the
sweeping up of the ambient medium (assuming a fraction $\epsilon_{\rm
ent}$ is swept up), and the thermal pressure of the ambient medium.
Metals are distributed within the solid angle $4\pi/N$ around the
angle in question and within the radius at which the test shell stalls
(i.e. has small velocity with respect to the ambient medium).

\begin{figure}
\plotone{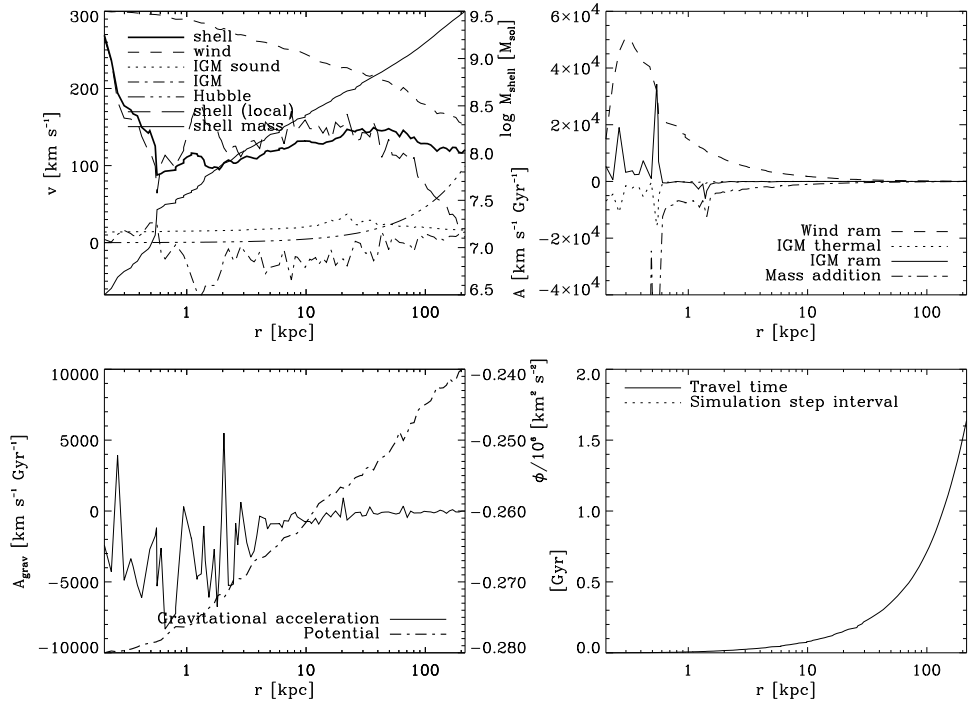}
\caption{ Sample of shell propagation starting at $z=4$, for a shell
with initial velocity of $\sim 300\kms$ at initial radius $200\,$pc in
a galaxy of baryonic mass $1.2\times 10^{9}\msol$.  {\bf Panel 1:}
physical radial velocities (with respect to the galaxy center where
appropriate) of the shell, the outflowing wind, the Hubble flow, and
the IGM.  We give also the local sound speed of the IGM, and the shell
velocity in the frame of the ambient gas, as well as the mass of the
shell (right axis).  {\bf Panel 2:} acceleration $(1/m_{\rm
shell})dp_{\rm shell}/dt$ of the shell due to the ram pressure of the
IGM, the ram pressure of the wind, the thermal pressure of the IGM,
and the acceleration due to the addition of mass to the shell (i.e.\
the term $(v/m)(dm/dt)$ where $v$ and $m$ are the velocity and mass of
the shell.  Note that this can slow down the shell even if the IGM ram
pressure adds momentum).  {\bf Panel 3:} Acceleration due to gravity
(left axis) and gravitational potential (right axis). {\bf Panel 4:}
Elapsed time since launch at initial radius.
\label{fig-intsampz3}
}
\end{figure}

	To study IGM enrichment by galaxies at $z \ga 3$ we have
applied our method to an SPH simulation (see Weinberg et al. 1999)
with $128^3$ gas and $128^3$ dark particles in a 17\,Mpc box.  This
simulation resolves galaxies of baryonic mass $\ga 3\times 10^8\msol$,
and therefore captures the bulk the forming stellar mass at $z \la 6$
(enrichment at higher-$z$ is not captured well).  Figure 1 gives an
example of the shell propagation in one galaxy at $z\approx 4$, and
shows that the ram pressure of the wind, the sweeping up of matter,
and the galaxy's gravity dominate the shell dynamics.  In this case,
the shell stalls about 200 (physical) kpc from the galaxy center after
about 1.5\,Gyr.  Figure 2 shows how far the shells propagate in a
large sample of $z \approx 4$ galaxies assuming $v_{\rm out}=300\kms$
and $\epsilon_{\rm ent}=0.1$; we find that these winds do escape to
large radii ($\ga 100\,$kpc) for galaxies of baryonic mass $\la
5\times10^9\msol$, limited chiefly by the available time between wind
launch and the time corresponding to the redshift at which the
enrichment is observed.  Our results are in accord with some some
previous analytical and numerical work, as shown in panel D.  There we
plot the supernova energy generation rate of the simulated galaxies,
and indicate (in the vertically shaded region) the corresponding SFR
range found by Pettini et al. (2001).  The dashed and dot-dashed lines
show the critical wind-escape luminosities of Silich \& Tenorio-Tagle
(2001) for spherical and disk galaxies, respectively; and the
simulated galaxies exceed both.  The lower shaded region shows the
parameter space probed by Mac Low \& Ferrara (1999), who found that
winds could escape with large entrainment (`blow-away') only from
$\sim 10^6\msol$ galaxies -- but note that those are the only ones for
which their assumed supernova luminosites are comparable to those of
the simulated galaxies.

\begin{figure}[tpb]
\plotone{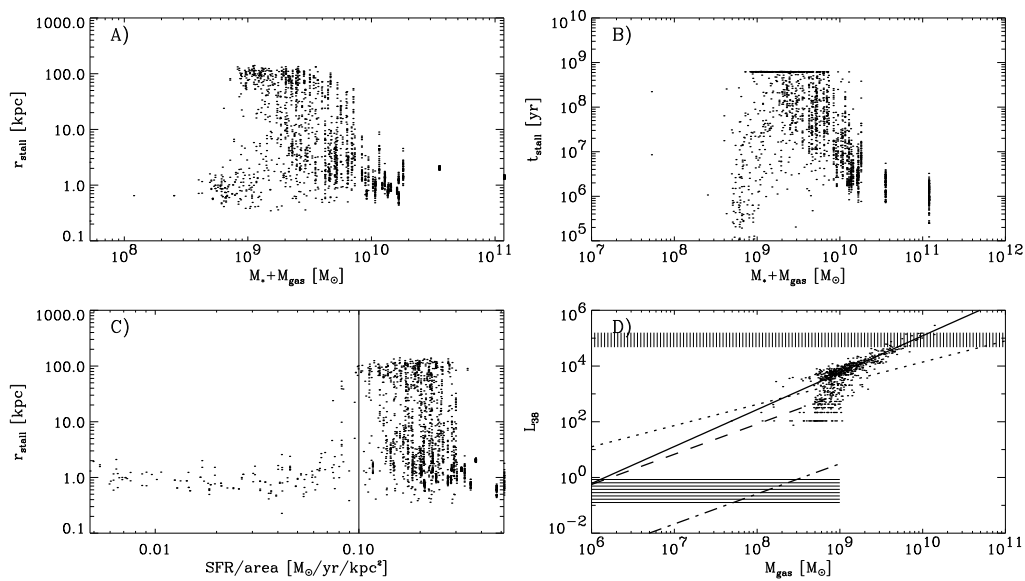}
\caption{ Quantities at wind-driving galaxies launching winds at
$z=4$. {\bf Panel A:} Wind stopping radius $r_{\rm stall}$ (or radius
at $z=3$ if smaller) versus galactic baryonic mass.  {\bf B:} Time
between shell launch at $z=4$ and stalling or observation at $z=3$.
{\bf C:} $r_{\rm stall}$ vs. total SFR/area of galaxy.  {\bf D:}
Supernova energy generation rate in units of
$10^{38}{\rm\,erg\,s^{-1}}$ versus galactic baryonic mass (see text
for more information).
\label{fig-wqz3}}
\end{figure}

	The enrichment of the IGM resulting from the winds is shown in
Figure 3 for various assumptions about $v_{\rm out}$ and
$\epsilon_{\rm ent}.$ The figure reveals that the enrichment of
low-density regions is quite sensitive to these parameters, but that
for some values the winds are able to enrich most gas particles to
metallicities comparable to those observed in the Ly$\alpha$ systems.
While the metallicity of simulation gas particles cannot be directly
compared to that of absorption-line systems (we are currently
implementing a procedure to generate simulated spectra from the
simulations), our results indicate that winds at $z \la 6$ can in
principle enrich the IGM to the observed level, given reasonable wind
parameters.

\begin{figure}
\plotone{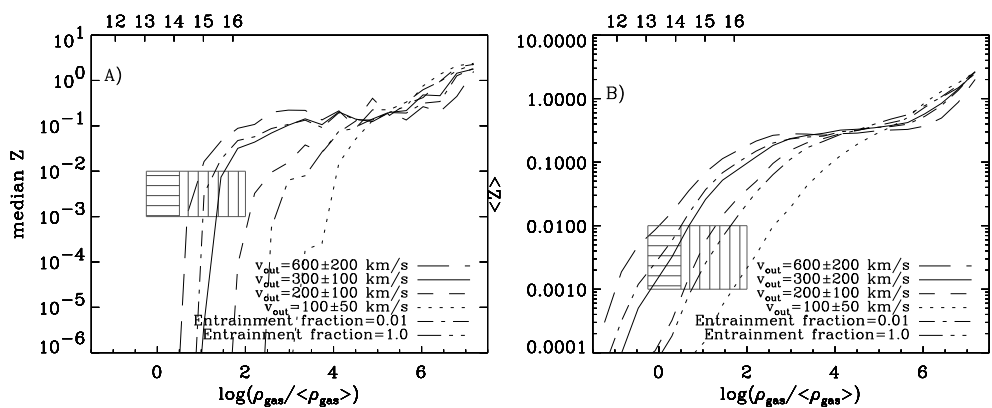}
\caption{ {\bf Panel A:} Median particle metallicity versus
$\delta\equiv \rho_{\rm gas}/\langle\rho_{\rm gas}\rangle$ for wind
models with $v_{\rm out}=100,200,300,600\kms$, and for $v_{\rm
out}=300\kms$ with $\epsilon_{\rm ent}=0.01$ or $\epsilon_{\rm
ent}=1$.  The top axis (here and in all panels) gives approximate
$\log N(H\,I)$ for an absorber of the given overdensity, using the
relation of Schaye (2001). The shaded box roughly indicates the
metallicity of low-column density Ly$\alpha$ clouds. {\bf B:} As for
panel A, but mass-weighted {\em mean} metallicities are plotted.
\label{fig-windres}
}
\end{figure}

Another interesting result of our calculations concerns the winds in
Lyman-break galaxies observed by Pettini et al. (2001).  These winds
have $v_{\rm out} \sim 250-1250\kms$ and mass outflow rates comparable
to their SFRs (i.e. $\chi\sim 1$).  Using these parameters, we find
that even with $\epsilon_{\rm ent}=1$ (i.e. the winds snowplough {\em
all} of the gas in their path), winds lauched at $z\sim 4$ can escape
to $\sim 200\,$kpc by $z\sim 3$ (see Figure 4)-- there is simply too
much momentum contained in these winds for any of the relevant forces
to confine them.  This prediction may, in fact, be borne out by
observations of a significant flux decrement in absorption spectra
within $\sim 200\,$kpc of Lyman-break galaxies (Pettini, this volume).

\section{Conclusions and Questions}

	We have developed a fast and flexible method of investigating
the enrichment of the IGM by various mechanisms, using numerical
simulations.  Applying this method to winds in simulations at $z \ga
3$, we find a few general results: 1) Essentially all galaxies at $z
\ga 3$ have specific SFRs as high as those of nearby wind-driving
galaxies. 2) Winds in low mass ($M \la 10^{10}\msol$) galaxies at $3
\la z\la 6$ with outflow velocites comparable to those in nearby
galaxies or as observed in Lyman-break galaxies should escape to large
radii. 3) The resulting pollution of the IGM is sufficient to roughly
account for the quantity of metals in the observed Ly$\alpha$ forest. 4)
In general, wind propagation is sensitive to the wind outflow velocity
and entrainment fraction.  Gravity and the assumed outflow velocity
determine the mass range of galaxies from which winds can escape, and
the distance to which they propagate is limited primarily by the
available time.

Our study so far has delineated a number of important theoretical and
observational questions.  Among them: 1) How uniform is the observed
enrichment of the IGM?  Are there pristine regions anywhere? 2) How
far are metals from galaxies? 3) How important is metal enrichment at
$z \gg 6$?  Would this metal enrichment be more uniform or less
uniform than that at lower redshift? 4) If winds enrich the IGM, do
they overly-disturb its properties, spoiling the agreement between
simulations and observations of the Ly$\alpha$ forest? 5) What
information about feedback during galaxy formation can be recovered
from the observed enrichment of the IGM?

\begin{figure}
\plotone{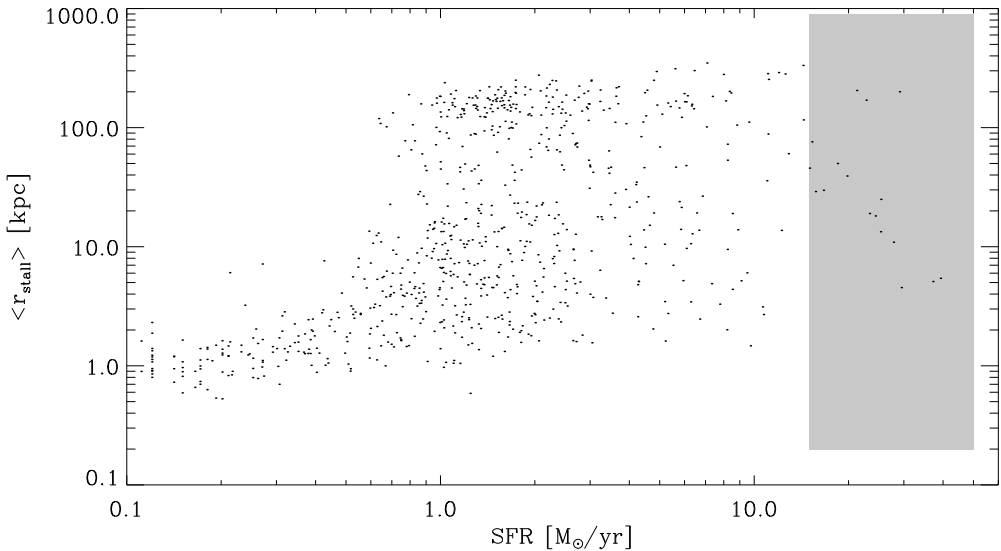}
\caption{ Maximum propagation (until $z=3$) radius in each galaxy
vs. star formation rate, for winds lauched at $z=4$ with $v_{\rm out}
= 750\pm500\kms$ and entrainment fraction unity (wind escape with a
more realistic entrainment traction would escape quite easily). The
Lyman-break galaxies of Pettini et al. (2001) show SFRs of
$10-50\msol\,{\rm yr^{-1}}$.
\label{fig-lybr}}
\end{figure}

The answers to these questions will shed considerable light on the
history of galaxy formation, and on the impact of galaxies on the
cosmic medium from which they form.

\end{document}